\documentclass[letterpaper,12pt]{article} 
\usepackage{amsmath,amssymb}

\providecommand{\e}{\ensuremath{\mathrm{e}}}
\providecommand{\hth}{\ensuremath{\hat\theta}}
\providecommand{\slD}{{\slash\!\!\!\!D}}

\author{W.~M\"uck\thanks{Email address: \texttt{wmueck@sfu.ca}}\
~and K.~S.~Viswanathan\thanks{Email address: \texttt{kviswana@sfu.ca}}\\
\emph{\small Department of Physics, Simon Fraser University, Burnaby,
British Columbia, V5A1S6 Canada}}
\title{The Wess-Zumino Model and\\ the AdS$_4$/CFT$_3$
Correspondence}

\begin{document}
\maketitle
\begin{abstract}
We consider the non-interacting massive Wess-Zumino model in four-dimensional
anti-de Sitter space and show that the conformal dimensions of the
corresponding boundary fields satisfy the relations expected from
superconformal invariance. In some cases the irregular mode
must be used for one of the scalar fields.
\end{abstract}

\section{Introduction}
\label{intro}
The AdS/CFT correspondence, originally conjectured by Maldacena
\cite{Maldacena} formulates a duality between a field theory on anti-de
Sitter space (AdS) and a conformal field theory (CFT) on its
boundary. The most noted example is the duality between AdS type IIB 
string theory and $\mathcal{N}=4$ super Yang-Mills theory
\cite{Girardello}. The precise form of the AdS/CFT
correspondence \cite{Gubser,Witten} in classical approximation reads
  \[ \exp \left(-I_{AdS}[\phi]\right) = \left\langle \exp \left(\int
  d^d x \phi_0(x) \mathcal{O}(x)\right)\right\rangle. \]

On the AdS side, the function $\phi_0$ represents the boundary value of the
field $\phi$, whereas on the CFT side it couples as a current to the
conformal field operator $\mathcal{O}$. There is a characteristic
relation between the mass of the AdS field $\phi$ and the conformal
dimension of the CFT field $\mathcal{O}$ \cite{Ferrara3}. 
This has been investigated for scalar \cite{Witten,Mueck1,Freedman}, spinor
\cite{Henningson,Mueck2}, vector \cite{Freedman,Mueck2,Minces},
graviton \cite{Liu,Arutyunov,Mueck3} and Rarita-Schwinger fields
\cite{Corley,Volovich,Koshelev}. Supersymmetry and supergravity in the
AdS/CFT context have been considered in
\cite{Ferrara1,Ferrara2,Banados,Ferrara3,Nishimura,Gunaydin,Ito,Anselmi,Figueroa,Yamaguchi}. 

For AdS supersymmetric field theories,
supersymmetry relates the masses of fields in the same
multiplet with each other. Hence, the AdS/CFT correspondence predicts that the
conformal dimensions of the corresponding boundary CFT operators
satisfy specific relations. On the other hand, superconformal symmetry
imposes a condition on the conformal dimensions of the primary fields of a
superconformal multiplet. One would expect that the AdS prediction
coincides with the CFT condition, which would mean that the AdS/CFT
correspondence couples an AdS super
multiplet to a superconformal multiplet on the AdS boundary. However,
to the best of our knowledge, no direct comparison has yet been made. 

In this paper, we shall tackle this problem by looking at the
non-interacting massive Wess-Zumino model in AdS$_4$, finding the
relations between the conformal dimensions of the corresponding scalar
and spinor boundary operators and comparing them with the relations
expected from superconformal invariance. We shall find in agreement
with the classic AdS papers \cite{Breitenlohner,Burges} that in some
cases one must use the irregular mode for one of the AdS scalar fields
in order for the AdS/CFT correspondence to hold true. This modifies 
the standard prescription \cite{Witten}, which uses only the
regular modes. 

Let us start with some preliminaries and use them to explain our
notation. For simplicity, we shall consider AdS$_4$ with Euclidean
signature. As is well known \cite{Weinberg}, it can be constructed as
a hyperboloid embedded into a five-dimensional Minkowski space with
a metric tensor $\eta_{AB}$ ($A,B=-1,0,1,2,3$), where 
\begin{equation}
\label{symm:minkmet}
  \eta_{-1-1}=-1, \quad \eta_{\mu\nu}= \delta_{\mu\nu}, \quad \text{and} 
  \quad \eta_{-1\mu}=0
\end{equation}  
($\mu,\nu=0,1,2,3$). Then, AdS$_4$ can be  defined by the embedding 
\begin{equation}
\label{symm:embed}
  y^A y^B \eta_{AB} = -1, \qquad y^{-1}>0,
\end{equation}
where the ``radius'' of the hyperboloid has been
chosen equal to 1 for simplicity. The metric  
\begin{equation}
\label{symm:metr1}
  ds^2 = dy^A dy^B \eta_{AB}
\end{equation}
represents the AdS metric, if one takes the $y^\mu$ as AdS$_4$
coordinates and defines $y^{-1}$ via equation \eqref{symm:embed}.  

While the representation \eqref{symm:embed} proves useful for finding
the AdS symmetries, a change of variables will reveal the conformal
symmetry of the AdS boundary. Introducing the variables $x^\mu$ by  
\begin{equation}
\label{symm:x}
  x^0 = \frac1{y^0 + y^{-1}}, \qquad x^i = x^0 y^i \quad (i=1,2,3),
\end{equation}
yields a representation of AdS$_4$ as the upper half space
$0<x^0<\infty$, $x^i\in\mathbb{R}$ with the metric  
\begin{equation}
\label{symm:xmet}
  ds^2 = \frac{\delta_{\mu\nu}}{(x^0)^2} dx^\mu dx^\nu.
\end{equation}

The use of the Minkowski five-space suggests the introduction of  
$4\times4$ gamma matrices $\hat\gamma_A$ satisfying
$\{\hat\gamma_A,\hat\gamma_B\}=2\eta_{AB}$. The spin matrices of the
corresponding Lorentz algebra in five dimensions are $\hat S_{AB}
=\frac14[\hat\gamma_A,\hat\gamma_B]$. 
The gamma matrices of the four-dimensional Euclidean Lorentz frame of
AdS$_4$ are given by 
\begin{equation}
\label{intro:gamma4}
  \gamma_a = \hat\gamma_a \hat\gamma_{-1}
\end{equation}
satisfying $\{\gamma_a,\gamma_b\}=2\delta_{ab}$,
($a,b=0,1,2,3$). The corresponding spin matrices are $S_{ab}
=\frac14[\gamma_a,\gamma_b]$. Covariant gamma matrices are defined by
$\Gamma_\mu=\e^a_\mu \gamma_a$ and covariant spin matrices by
$\Sigma_{\mu\nu}=\e^a_\mu e^b_\nu S_{ab}$.  

Finally, let us give a short outline of the rest of the paper. The
AdS$_4$ symmetry algebra and its $\mathcal{N}=1$ grading will be
derived in section~\ref{symm}. In section~\ref{conf} we recast these
algebras in the form of conformal and superconformal algebras and
recall the relations between the conformal weights of the primary fields in a
superconformal multiplet. The AdS$_4$ superspace is
constructed in section~\ref{sup}. In section~\ref{mod} we consider the
Wess-Zumino model in AdS$_4$ and calculate the conformal dimensions of
the corresponding boundary fields. We refer our readers to the appendices
\ref{app:grass} and \ref{app:calc} for information on Grassmann
variables and the calculation of Killing spinors, respectively. 

\section{Symmetry Algebra and its $\mathcal{N}=1$ Grading}
\label{symm}
The AdS Symmetries are easiest found considering the embedding
\eqref{symm:embed}. In fact, equation \eqref{symm:embed} is invariant
under Lorentz transformations of the Minkowski five-space, which are of
the form $(y')^A = \mathcal{R}^A_{\;B} y^B$, where the matrix
$\mathcal{R}$ satisfies $\mathcal{R}^T\eta \mathcal{R} =\eta$ and
$\mathcal{R}^{-1}_{\;-1}>0$. For the purposes of this paper we
consider only the connected subgroup of such matrices, namely the Lie
group $SO(4,1)$. An infinitesimal transformation, $\mathcal{R} =1+M$,
has the form 
\begin{equation}
\label{symm:ysymm}
  \delta y^A = M^A_{\;B} y^B = \frac12 \omega^{CD} (M_{CD})^A_{\;B} y^B,
\end{equation} 
where the generators
$(M_{CD})^A_{\;B}=\delta^A_C\eta_{BD}-\delta^A_D\eta_{BC}$ form the
standard basis of the $so(4,1)$ algebra and satisfy the commutation
relations  
\begin{equation}
\label{symm:malg}
  \left[ M_{AB},M_{CD} \right] = \eta_{AD} M_{BC} +\eta_{BC} M_{AD}
  - \eta_{AC} M_{BD} -\eta_{BD} M_{AC}.
\end{equation}

One can show from the equations \eqref{symm:x} and \eqref{symm:ysymm}
that the infinitesimal change of the coordinates $x^\mu$ is given by 
\begin{equation}
\label{symm:xsymm}
\begin{aligned}
  \delta x^0 &= - x^0 (\lambda+2 a_i x^i),\\
  \delta x^i &= - x^i (\lambda+2 a_j x^j) +a^i x^2 -b^i +\omega^{ij} x^j,
\end{aligned}
\end{equation}
where $x^2=x^\mu x^\nu \delta_{\mu\nu}$ and where the parameters
$a^i$, $b^i$ and $\lambda$ are defined by
\begin{equation}
\label{symm:pardef}
  a^i=\frac12 \left(\omega^{0i}+\omega^{-1i}\right),
  \quad b^i=\frac12\left(\omega^{0i}-\omega^{-1i}\right)
  \quad \text{and} \quad \lambda= \omega^{-10},
\end{equation}
respectively. Obviously, the transformations \eqref{symm:xsymm} reduce
to infinitesimal conformal transformations on the boundary, $x^0=0$. 

It is straightforward to find the (complex) $\mathcal{N}=1$ grading of the
$so(4,1)$ algebra \eqref{symm:malg}. First, introduce the fermionic
generators $Q^\alpha$ ($\alpha=1,2,3,4$), which transform as $so(4,1)$
spinors, i.e.\  
\begin{equation}
\label{symm:mq1}
  \left[ M_{AB}, Q^\alpha \right] = - (\hat S_{AB})^\alpha_{\;\beta} Q^\beta.
\end{equation}
Then, the superalgebra closes with the anti-commutator (see
appendix~\ref{app:grass} for notation)  
\begin{equation}
\label{symm:qq1}
  \left\{Q^\alpha, Q^\beta\right\} = -2(\hat S^{AB}\hat C^{-1})^{\alpha\beta}
  M_{AB}.
\end{equation}
We would like to add two remarks at this point. First, the validity of
equation \eqref{symm:qq1} is conditional upon the fact that we grade
the five-dimensional Minkowski algebra. For higher dimensions (e.g.\
AdS$_5$) one would have to introduce additional bosonic operators to
obtain closure of all Jacobi identities \cite{vanHolten}.  
Second, the equations \eqref{symm:malg}, \eqref{symm:mq1} and \eqref{symm:qq1} 
define the complex superalgebra $B(0/2)$, whose real form is
$osp(1,4)$ \cite{Cornwell3}. Unfortunately, $osp(1,4)$ does not
contain $so(4,1)$ in its even part, which means in other words that no
Majorana spinors exist for our Minkowski five-space. However, $osp(1,4)$
contains $so(3,2)$, which is the symmetry group of AdS$_4$ with
Minkowski signature. Resorting to a Wick rotation at the end to make
our results valid, we shall ignore this fact and formally carry out
the analysis.  

\section{Conformal and Superconformal Algebra}
\label{conf}
As mentioned in section~\ref{symm}, the AdS symmetry group acts as the
conformal group on the AdS boundary. In this section, we shall for completeness
explicitely show the isomorphisms between the $d=3$ conformal algebra and
$so(4,1)$ as well as between their $\mathcal{N}=1$ superalgebras. 
Let us introduce the \emph{conformal basis} of $so(4,1)$ by defining
\begin{equation}
\label{conf:bas}
\begin{aligned}
  D &= M_{-10}    ,&\qquad  K_i &=M_{0i}+M_{-1i},\\
  L_{ij} &= M_{ij},&        P_i &=M_{0i}-M_{-1i}.
\end{aligned}
\end{equation} 
Then, an element $M\in so(4,1)$ takes the form 
\begin{equation}
\label{conf:algelem}
  M = \frac12 \omega^{AB} M_{AB} = \lambda D + a^i K_i + b^i P_i
  +\frac12 \omega^{ij} L_{ij},
\end{equation}
with the parameters $a_i$, $b_i$ and $\lambda$ given by equation
\eqref{symm:pardef}.  
One easily finds from equation \eqref{conf:bas} the commutation
relations of $D$, $P_i$, $K_i$ and $L_{ij}$, which are given by
\begin{equation}
\label{conf:calg}
\begin{aligned}
  \left[ D,P_i\right] &= -P_i,\\
  \left[ D,K_i\right] &= K_i,\\
  \left[ L_{ij},P_k\right] &= \delta_{jk}P_i-\delta_{ik}P_j,\\
  \left[ L_{ij},K_k\right] &= \delta_{jk}K_i-\delta_{ik}K_j,\\
  \left[ P_i,K_j\right] &= 2 \left(\delta_{ij}D -L_{ij}\right),\\
  \left[ L_{ij},L_{kl}\right] &= \delta_{il}L_{jk}+\delta_{jk}L_{il}
  -\delta_{ik}L_{jl} - \delta_{jl}L_{ik},\\
  \left[ P_i,P_j\right] &=  \left[ K_i,K_j\right] =
  \left[ D,L_{ij}\right]=0.
\end{aligned}
\end{equation}
Equations \eqref{conf:calg} are the standard representation of the
conformal algebra \cite{diFrancesco}.

The $\mathcal{N}=1$ grading of the conformal algebra \eqref{conf:calg}
is well known in the literature \cite{vanHolten}, but again, a direct
comparison with the superalgebra given in section~\ref{symm} seems
useful. This is done by choosing a particular representation of the
five-dimensional Clifford algebra of matrices $\hat\gamma_A$. Choosing
\begin{equation}
\label{conf:gammas}
  \hat\gamma_i = \begin{pmatrix} \sigma_i & 0 \\
				 0 & -\sigma_i \end{pmatrix}, \quad
  \hat\gamma_0 = \begin{pmatrix} 0 & \mathbf{1} \\
				 \mathbf{1} & 0 \end{pmatrix}, \quad
  \hat\gamma_{-1} = \begin{pmatrix} 0 & \mathbf{1} \\
				 -\mathbf{1} & 0 \end{pmatrix}, 
\end{equation}
where $\sigma_i$ are the Pauli spin matrices and $\mathbf{1}$ is the
$2\times2$ unit matrix, one easily finds from the definition
\eqref{conf:bas} the spinor representations of the conformal basis
elements, which are 
\begin{equation}
\label{conf:cspin}
\begin{aligned}
  \hat S(L_{ij}) &= \frac12 \begin{pmatrix} \sigma_{ij} & 0 \\
				0 & \sigma_{ij} \end{pmatrix}, \quad &
  \hat S(D) &= \frac12 \begin{pmatrix} \mathbf{1} & 0 \\
				 0 & -\mathbf{1} \end{pmatrix},\\ 
  \hat S(P_i) &= \begin{pmatrix} 0 & 0 \\
				\sigma_i & 0 \end{pmatrix}, &
  \hat S(K_i) &= \begin{pmatrix} 0 & -\sigma_i \\
				0 & 0 \end{pmatrix}.
\end{aligned}
\end{equation}
Splitting the spinor operator $Q^\alpha$ into two 2-component spinors,
  \[ Q^\alpha = \begin{pmatrix} q\\ s \end{pmatrix}, \]
we find from equation \eqref{symm:mq1} the commutators
\begin{equation}
\label{conf:cqs}
\begin{aligned}
  {}[L_{ij},q^\alpha] &= -\frac12 (\sigma_{ij} q)^\alpha, \quad&
  [L_{ij},s^\alpha] &= -\frac12 (\sigma_{ij} s)^\alpha, \\ 
  [D,q^\alpha] &= -\frac12 q^\alpha, & [D,s^\alpha] &=\frac12 s^\alpha, \\
  [P_i,q^\alpha] &=0, &  [P_i,s^\alpha] &= -(\sigma_i q)^\alpha, \\
  [K_i,q^\alpha] &= (\sigma_i s)^\alpha, & [K_i,s^\alpha] &=0.
\end{aligned}
\end{equation} 
Furthermore, the charge conjugation matrix $\hat C$ has the form 
\[ \hat C = \begin{pmatrix} 0 & c \\ c & 0 \end{pmatrix}, \]
where $c$ is the charge conjugation matrix in three dimensions. 
Hence, using the identity 
\begin{equation}
\label{conf:ident}
  \hat S^{AB} M_{AB} = -2 \hat S(D) D + \hat S(K_i) P_i 
  +\hat S(P_i) K_i + \hat S(L_{ij}) L_{ij},    
\end{equation}
equations \eqref{symm:qq1} and \eqref{conf:cspin} yield the
anticommutators
\begin{equation}
\label{conf:acomm}
\begin{aligned}
  \{q^\alpha,q^\beta\} &=2\left(\sigma_i c^{-1}\right)^{\alpha\beta}P_i,\\
  \{s^\alpha,s^\beta\} &=-2\left(\sigma_i
  c^{-1}\right)^{\alpha\beta}K_i,\\
  \{q^\alpha,s^\beta\} &= \left(2c^{-1}D - 
  \sigma_{ij} L_{ij}\right)^{\alpha\beta}.
\end{aligned}
\end{equation}
Equations \eqref{conf:calg}, \eqref{conf:cqs} and \eqref{conf:acomm}
form the $\mathcal{N}=1$ superconformal algebra in three dimensions
\cite{vanHolten}. 

Obviously, the operators $L_{ij}$, $P_i$ and $q^\alpha$ form the 
three-dimensional $\mathcal{N}=1$ Poincar\'e superalgebra. Let us
therefore consider a scalar super-Poincar\'e multiplet consisting of
the scalar fields $\mathcal{O}$ and $\mathcal{F}$ and the spinor
field $\chi$, which satisfy the supersymmetry relation
\begin{equation}
\label{conf:mult}
\begin{aligned}
  q^\alpha \mathcal{O} &= \chi^\alpha, \\
  q^\alpha \chi^\beta &= (\sigma_i c^{-1})^{\alpha\beta} \partial_i
  \mathcal{O} + (c^{-1})^{\alpha\beta} \mathcal{F}, \\
  q^\alpha \mathcal{F} &= -(\sigma_i \partial_i \chi)^\alpha.
\end{aligned}
\end{equation}
Imposing conformal symmetry on the multiplet means that
the scaling dimensions of the fields must satisfy
\begin{equation}
\label{conf:drel}
  \Delta_\mathcal{O} = \Delta_\chi +\frac12 = \Delta_\mathcal{F}+1.
\end{equation}
This relation is obtained by acting with the commutator $[D,q]$ on the
fields $\mathcal{O}$ and $\chi$. Notice that the spinor operator $s$
is expressed in terms of $K_i$ and $q$ and thus does not introduce
new fields into the multiplet.  

\section{Construction of AdS Superspace}
\label{sup}
In order to obtain the ($\mathcal{N}=1$) supersymmetric extension of AdS$_4$,
one first introduces Grassmann coordinates $\hth^\alpha$ in addition
to the AdS coordinates $x^\mu$. Then one postulates that the symmetry algebra
of the superspace is given by the graded Lie algebra constructed in
section~\ref{symm}. Because the symmetry algebra determines
infinitesimal coordinate transformations, one can determine the latter from the
knowledge of the algebra. The method to be used has been described by Zumino
\cite{Zumino} and applies to any group $\mathcal{G}$ with a subgroup
$\mathcal{H}$. 

In the case at hand, an element $g\in \mathcal{G}$ is uniquely represented by
\begin{equation}
\label{sup:g}
  g = \e^{\hat\xi Q} h(x),
\end{equation}
where $\hat\xi$ is some Grassmann coordinate spinor, whose relation with
$\hth$ will be defined later and $h(x)\in\mathcal{H}=SO(4,1)$ is a
Lie-algebra valued function of the coordinates $x^\mu$. Then, by virtue of the
group axioms, one can write
\begin{equation}
\label{sup:gg}
  g_0 g = g' = \e^{\hat{\xi'}Q} h(x')
\end{equation}
and consider the transformations $\hat\xi\rightarrow\hat\xi'$ and
$x\rightarrow x'$ as 
induced by the group element $g_0$. We shall in the following use the
abbreviations $\Theta=\hat\xi Q$ and $M=\frac12\omega^{AB}M_{AB}.$
In the case of $g_0\in \mathcal{H}$, i.e.\ an even transformation, equation
\eqref{sup:gg} takes the form
$\e^M \e^\Theta h(x)=\e^{\Theta'}\e^{M'} h(x)$,
where $M'$ and $\Theta'$ are determined by the Baker-Campbell-Hausdorff
formula. For infinitesimal $M$ one finds $M'=M$ and
$\Theta'=\Theta+[M,\Theta]$. By definition, the even part $\e^M h(x) = h(x')$ 
yields equation \eqref{symm:xsymm} and thus does not contain new information,
whereas the odd part yields a linear transformation of the Grassmann
coordinates, namely
\begin{equation}
\label{sup:xisymm}
  \delta\hat\xi^\alpha = \frac12 \omega^{AB} (\hat S_{AB})^\alpha_{\;\beta}
  \hat\xi^\beta.
\end{equation}

On the other hand, for $g_0=\e^R$ with the abbreviation $R=\hat\varepsilon Q$,
one writes
\begin{equation}
\label{sup:R}
  \e^R \e^\Theta h(x) = \e^{\Theta'} \e^M h(x).
\end{equation}
Then, for infinitesimal $R$ one finds \cite{Zumino} 
\begin{align}
\label{sup:dTheta}
  \delta\Theta &=\left(1+\frac13\Theta^2-\frac1{45}\Theta^4\right) \wedge R.\\
\intertext{and} \label{sup:M}
  M &=\left(-\frac\Theta2 +\frac{\Theta^3}{24} \right)\wedge R,
\end{align}
where the  notation
  \[ 1\wedge Y =Y, \quad X\wedge Y =[X,Y], \quad X^2 \wedge Y =[X,[X,Y]],
  \quad\text{etc.} \] 
has been used. Equations \eqref{sup:dTheta} and \eqref{sup:M} are evaluated
explicitly using the 
anti-commutator \eqref{symm:qq1} and various Fierz identities listed in
appendix~\ref{app:grass}, leading to
\begin{align}
\label{sup:dxi}
  \delta\hat\xi^\alpha &= \hat\varepsilon^\alpha \left[1-\frac56\hat\xi\hat\xi
  - \frac19(\hat\xi\hat\xi)^2 \right] -\frac16
  (\hat\gamma^A \hat\varepsilon)^\alpha (\hat\xi\hat\gamma_A\hat\xi)\\
\intertext{and} \label{sup:M2}
  M &= -\left( 1+\frac16 \hat\xi\hat\xi \right)
  (\hat\xi \hat S^{AB}\hat\varepsilon) M_{AB},
\end{align}
respectively. The transformation formula \eqref{sup:dxi} can be simplified by
defining
\begin{equation}
\label{sup:xidef}
  \hth = \hat\xi \left( 1-\frac13 \hat\xi\hat\xi\right).
\end{equation}
While $\hth$ still is an
$SO(4,1)$ spinor, i.e.\ it transforms under even transformations as 
\begin{equation}
\label{sup:thetasymm}
  \delta\hth^\alpha = \frac12 \omega^{AB}
  (\hat S_{AB})^\alpha_{\;\beta} \hth^\beta,
\end{equation}
the odd transformation laws, equations \eqref{sup:dxi} and
\eqref{sup:M2}, become
\begin{align}
\label{sup:dtheta}
  \delta\hth^\alpha &= \hat\varepsilon^\alpha
  \left[1-\hth\hth - \frac12
  (\hth\hth)^2\right],\\
\intertext{and} \label{sup:M3}
   M &= -\left( 1+\frac12 \hth\hth \right)
  (\hth \hat S^{AB}\hat\varepsilon) M_{AB},
\end{align}
respectively. Using equations \eqref{conf:algelem},
\eqref{conf:ident} and \eqref{intro:gamma4} one finds 
\begin{equation}
\label{sup:abc}
\begin{aligned}
  \lambda &= - \left(1+\frac12\hth\hth\right)
  \left(\hth \gamma_0 \hat\varepsilon \right), \\
  a^i &= -\frac12 \left(1+\frac12\hth\hth\right)
  \left[ \hth \gamma^i(1-\gamma_0) \hat\varepsilon \right],\\
  b^i &= \frac12 \left(1+\frac12\hth\hth\right)
  \left[ \hth \gamma^i(1+\gamma_0) \hat\varepsilon \right],\\
  \omega^{ij} &= -2 \left(1+\frac12\hth\hth\right)
  \left(\hth S^{ij} \hat\varepsilon \right).
\end{aligned}
\end{equation}

Thus, the supersymmetry transformation $\delta x^\mu$ is 
given by equation \eqref{symm:xsymm} using the parameters of equation
\eqref{sup:abc}. Although this solves the problem of finding the
superspace transformations, a space-time covariant formulation would be
much more desirable. Such a formulation involves the Killing spinors,
which are calculated in appendix~\ref{app:calc}. In fact, it is easy
to show from equation \eqref{app:kill} that the quantity
$\hat\eta\Lambda\Gamma^\mu\Lambda^{-1}\hat\varepsilon$ is a Killing
vector. On the other hand, because also $\delta x^\mu$ is a Killing vector
and is linear in 
$\hat\varepsilon$, it must have exactly this form with $\hat\eta$ being a 
function of $\hth$ only. A direct comparison using equations
\eqref{symm:xsymm}, \eqref{sup:abc}, \eqref{app:lsol}, 
\eqref{app:linv} and \eqref{app:kill} shows that 
\begin{equation}
\label{sup:dx}
  \delta x^\mu = \left( 1+\frac12 \hth\hth \right)
  \hth \Lambda \Gamma^\mu \Lambda^{-1} \hat\varepsilon.
\end{equation}
Equations \eqref{sup:dtheta} and \eqref{sup:dx} represent the
supersymmetry transformation of the AdS superspace in a space-time
covariant form. It is with this form that one can hope to effectively
carry out the calculations involving superfields. Moreover, it will
allow our formal results to be carried over to the Lorentzian signature case,
where Majorana spinors exist. Notice that $\delta x^\mu$ in equation
\eqref{sup:dx} is generically complex for Euclidean signature, because
no Majorana spinors exist in this case.  

Let us conclude this section by finding the invariant integral measure for
integration over the AdS$_4$ superspace. First, we observe that the
bosonic part $d^4x \sqrt{g(x)} = d^4 x (x^0)^{-4}$ is in itself
invariant under any variable transformation, i.e.\ also under the
supersymmetry transformation \eqref{sup:dx}. For the fermionic part of
the integral measure, let us make the ansatz $d^4\hth
\rho(\hth)$ and demand that it be invariant under the
transformation $\hth\rightarrow\hth'=\hth+\delta\hth$,
where $\delta\hth$ is given by equation \eqref{sup:dtheta}. 
From equation \eqref{sup:dtheta} follows that
\begin{equation}
\label{sup:mea1}
  d^4\hth'= d^4\hth \left[1-2 \left(1+\hth\hth\right)
  \hth\hat\varepsilon\right].
\end{equation} 
Multiplying equation \eqref{sup:mea1} with $\rho(\hth')$ and
expanding to terms linear in $\hat\varepsilon$ we find the equation
  \[ \left[1-\hth\hth
  -\frac12(\hth\hth)^2\right] 
  \frac{\partial}{\partial\hth_\alpha} \rho = 2
  \hth^\alpha (1+\hth\hth) \rho, \]
whose solution up to a multiplicative constant is 
\begin{equation}
\label{sup:rho}
  \rho(\hth) = 1+ \hth\hth
  +\frac32(\hth\hth)^2.
\end{equation}
It is straightforward to show that $d^4\hth\rho(\hth)$ is
also invariant under the bosonic transformation \eqref{sup:thetasymm}.
Hence, the expression
\begin{equation}
\label{sup:measure}
  d^4x \sqrt{g(x)}\; d^4\hth \left[1+\hth\hth
  +\frac32(\hth\hth)^2 \right]
\end{equation}
is the invariant superspace integration measure.

\section{The Wess-Zumino Model}
\label{mod}
Let us start this section with the expansion of  a chiral superfield
in powers of the Grassmann variables $\hth$ in order to identify its
scalar and spinor field contents. Because of the existence of previous work
\cite{Keck,Ivanov,Breitenlohner,Burges} only the results will be
given. However, it should be noted that our derivation differs in some
points from \cite{Keck,Ivanov}. Keck \cite{Keck} coupled a spinor field
directly to the $SO(4,1)$ spinor variable $\hat\xi$ of
section~\ref{sup}, thereby demanding that the spinor field too be an
$SO(4,1)$ instead of a Lorentz spinor. On the other hand, Ivanov and
Sorin \cite{Ivanov} considered the Killing spinor $\theta$ (see
appendix~\ref{app:calc}) as the independent Grassmann variable, which
can directly be coupled to a Lorentz spinor field. However, the
complicated transformation rule for $\theta$ under supersymmetry
transformations is a minor drawback of their very complete
formulation, which led us to consider the $SO(4,1)$ spinor $\hth$ as
the independent superspace variable and realize the coupling to
Lorentz spinor fields via a matrix $\Lambda(x)$, which is calculated
in appendix~\ref{app:calc}. We feel that this treatment combines the
nice features of both references, \cite{Keck} and \cite{Ivanov}. In
addition, it yields the Killing spinor $\theta$ as a side product. 

The Wess-Zumino multiplet is given by the scalar fields $A$, $B$, $F$
and $G$ and the Dirac spinor field $\psi$. Their supersymmtry
transformations are easiest found by considering chiral superfields. 
Therefore, let us define 
\begin{equation}
\label{mod:abfg}
\begin{aligned} 
  A &= A_L + A_R, \qquad & B &=A_L -A_R, \\
  F &= F_L + F_R,        & G &=F_L -F_R,
\end{aligned}
\end{equation} 
and let us introduce the chiral projection operators 
\begin{equation}
\label{mod:LR}
  L=\frac12 (1-i\hat\gamma_{-1}) \quad \text{and} \quad 
  R=\frac12 (1+i\hat\gamma_{-1}).
\end{equation}
Then, the left and right handed chiral superfields are given by
\begin{align}
\notag 
  \Phi_L(x,\hth) &= A_L + \hth \Lambda L \psi +
  (\hth \Lambda L \Lambda^{-1} \hth) F_L \\ 
\label{mod:phiL}
  &\quad -\frac{i}2 (\hth\Lambda \hat\Gamma^\mu
  \Lambda^{-1}\hth) D_\mu A_L +\frac12 (\hth\hth)\hth\Lambda \slD L\psi
  +\frac18 (\hth\hth)^2 D^\mu D_\mu A_L, \\
\notag
  \Phi_R(x,\hth) &= A_R + \hth \Lambda R\psi +
  (\hth\Lambda R \Lambda^{-1}\hth) F_R \\ 
\label{mod:phiR}
  &\quad +\frac{i}2 (\hth\Lambda \hat\Gamma^\mu
  \Lambda^{-1}\hth) D_\mu A_R +\frac12 (\hth\hth)\hth\Lambda \slD L\psi
  +\frac18 (\hth\hth)^2 D^\mu D_\mu A_R, 
\end{align}
respectively. It is straightforward to show using the transformation rules
\eqref{sup:dtheta} and \eqref{sup:dx} that the supersymmetry
transformations of the left handed superfield components are given by
\begin{equation}
\label{mod:trans}
\begin{aligned}
  \delta A_L &= -\varepsilon L \psi,\\
  \delta (L\psi) &= - 2L (\slD A_L +F_L) \varepsilon,\\
  \delta F_L &= \varepsilon L\psi - \varepsilon \slD L \psi,
\end{aligned}
\end{equation}
where we introduced the Killing spinor $\varepsilon_\alpha =
(\hat\varepsilon \Lambda)_\alpha$.
It takes somewhat more effort to show that
all terms in the expansion \eqref{mod:phiL} transform correctly.
To find the supersymmetry transformations of $A_R$, $F_R$ and $R\psi$,
simply replace $L$ with $R$ in equation \eqref{mod:trans}.

For the Wess-Zumino model we also introduce ``conjugate'' superfields
by defining
\begin{equation}
\label{mod:conj}
\begin{aligned} 
  \bar \Phi_R &= \bar A_L + \theta R \bar\psi + (\theta R \theta) \bar
  F_L +\cdots,\\
  \bar \Phi_L &= \bar A_R + \theta L \bar\psi + (\theta L \theta) \bar
  F_R +\cdots ,
\end{aligned}
\end{equation}
where we used the Killing spinor $\theta_\alpha=(\hth\Lambda)_\alpha$
and where the dots indicate terms similar to those in equations
\eqref{mod:phiL} and \eqref{mod:phiR}.

A manifestly supersymmetric action is then given by the expression
\begin{equation}
\label{mod:action1}
  S = \int d^4x \sqrt{g(x)}\; 
  d^4\hth \rho(\hth) 
  \left[ \bar\Phi_L\Phi_R + \bar\Phi_R\Phi_L 
  - m \left( \bar\Phi_L\Phi_L +\bar\Phi_R\Phi_R \right)\right],
\end{equation}
which describes the non-interacting Wess-Zumino model with a mass
term. After inserting the integration measure \eqref{sup:rho} 
we can perform the Berezin integration and re-express the result in
terms of the fields $A$, $B$, $F$ and $G$. 
Hence, we obtain (up to a multiplicative
constant and surface terms, which have been dropped)
\begin{equation}
\label{mod:action2}
\begin{aligned}
  S_{bulk} &= \int d^4x \sqrt{g} \left[ \frac12\bar\psi
  \left(\slD-\overset{\leftarrow}{\slD}\right) \psi + D_\mu\bar A D^\mu A 
  +D_\mu\bar B D^\mu B \right. \\
  &\quad -3\bar A A -3\bar B B 
  -A\bar F -\bar A F -B\bar G -\bar B G -\bar F F -\bar G G \\
  &\left.\phantom{\frac12} - m \left(\bar\psi\psi 
  -3\bar A A + 3 \bar B B -A\bar F -\bar A F +B\bar G +\bar B G\right)\right].
\end{aligned}
\end{equation} 
Solving the equations of motion for the auxilliary fields $F$ and $G$ gives 
\begin{equation}
\label{mod:F}
  F = (m-1) A \quad \text{and} \quad G= -(m+1) B.
\end{equation}
Similar relations hold for $\bar F$ and $\bar G$.
Substituting equation \eqref{mod:F} back into the action
\eqref{mod:action2} yields the on-shell supersymmetric action
\cite{Breitenlohner,Burges}
\begin{equation}
\label{mod:action3}
\begin{aligned}
  S_{bulk} &= \int d^4x \sqrt{g} \left[ \frac12\bar\psi
  \left(\slD-\overset{\leftarrow}{\slD}\right) \psi - m \bar\psi \psi \right.\\
  &\quad + D_\mu\bar A D^\mu A + (m^2+m-2) \bar A A\\
  &\left.\phantom{\frac12} 
   + D_\mu\bar B D^\mu B + (m^2-m-2) \bar B B\right].
\end{aligned}
\end{equation}
The mass parameter $m$ describes the mass of the fermion
$\psi$. Moreover, for $m=0$ the scalar fields $A$ and $B$ are
conformally coupled. 
The bulk action $S_{bulk}$ has to be accompanied for the AdS/CFT
correspondence by a surface term derivable from the variational
principle \cite{Henneaux}. 

It seems straightforward to read off from equation \eqref{mod:action3}
the conformal dimensions of the
boundary operators coupling to the boundary values of $A$, $B$ and
$\psi$. However, care must be taken when specifying boundary
conditions. According to \cite{Breitenlohner,Burges}, because the fields
$A$, $B$ and $\psi$ are to form an irreducible representation of
$osp(1,4)$ (for Minkowski signature), one must use the irregular
mode for one of the scalar fields, if $|m|<\frac12$. 
Therefore, let us not exclude the irregular modes for the scalar fields.
Then the conformal dimensions of the boundary operators are given by 
\cite{Witten,Mueck1,Mueck2} 
\begin{equation}
\label{mod:deltas}
\begin{aligned}
  \Delta_A &= \frac32 \pm \left|\frac12+m \right|,\\
  \Delta_B &= \frac32 \pm \left|\frac12-m\right|,\\
  \Delta_\psi &= \frac32 +|m|,
\end{aligned}
\end{equation}
where the plus and minus signs correspond to using the regular and
irregular modes, respectively. Let us consider the case $m\ge0$. For
$m<0$ only the roles of $A$ and $B$ interchange.
Comparing the values \eqref{mod:deltas} with equation
\eqref{conf:drel} we can identify the boundary fields corresponding to
the AdS fields $A$, $B$, and $\psi$ with the primary conformal fields
$\mathcal{O}$, $\mathcal{F}$ and $\chi$, respectively. Moreover, if
$m<\frac12$, the irregular mode must be used for $B$ in order to make
this identification. 

In conclusion, we found for a simple example that the AdS/CFT
correspondence relates fields of AdS supersymmetry multiplets to the
primary fields of superconformal multiplets. 
This fact was derived by explicitly constructing the AdS
supersymmetric model and comparing its predictions with the relations
expected from super CFT. In some cases, irregular
modes must be considered for AdS fields, changing the standard
prescription of the AdS/CFT correspondence. This conclusion stems from
both, a pure AdS point of view and the AdS/CFT correspondence. 
We feel that a more general treatment should
be attempted in future work. 

\section*{Acknowledgements}
We are very grateful to D.\ Z.\ Freedman for pointing out to us the
importance of using the irregular modes, without which we would have drawn
the wrong conclusion. 

This work was supported in
part by a grant from NSERC. W.\ M.\ is grateful to Simon Fraser
University for financial support. 
 
\begin{appendix}
\numberwithin{equation}{section}
\section{Spinor Grassmann Variables}
\label{app:grass} 
This appendix summarizes our notations and useful
formulae for Grassmannian spinor variables in five dimensions. We shall
concentrate on facts which do not depend on the signature of the
five-dimensional metric, thus avoiding explicit matrix representations
and the introduction of complex conjugate spinors. Most of the
following is derived from information on Clifford algebras and
their representations, which can be found in \cite{Cornwell3}.

A spinor $\hth$ has components $\hth^\alpha$ ($\alpha=1,2,3,4$), which are
Grassmannian variables, i.e.\ the components of any two spinors $\hth$
and $\hat\eta$ satisfy
\begin{equation}
\label{app:acomm}
  \left\{\hth^\alpha,\hat\eta^\beta\right\} = 0.
\end{equation}
Spinor matrices usually carry an upper and a lower index, such as
$\delta^\alpha_\beta$, $(\hat\gamma_A)^\alpha_{\;\beta}$ etc. However,
indices can be lowered and raised with the charge conjugation matrix and its
inverse, respectively:
\begin{equation}
\label{app:ccon}
  \hth_\alpha = \hat C_{\alpha\beta} \hth^\beta, \qquad
  \hth^\alpha= (\hat C^{-1})^{\alpha\beta}\hth_\beta.
\end{equation}
For $D=5$ the charge conjugation matrix is anti-symmetric. One can now define
the scalar product of two spinors by
\begin{equation}
\label{app:scalprod}
  \hat\eta\hth = \hat\eta_\alpha\hth^\alpha
  = - \hat\eta^\alpha\hth_\alpha = \hth\hat\eta.
\end{equation}

The vector space of $4\times 4$ matrices with only lower indices is spanned by
16 matrices, which can conveniently be chosen to be
i) the anti-symmetric charge conjugation matrix $\hat C$,
ii) the 5 anti-symmetric matrices $(\hat C \hat\gamma_A)$ and
iii) the 10 symmetric matrices $(\hat C \hat S_{AB})$  \cite{Pilch}.
The symmetry properties of the latter two follow directly from 
\begin{align}
\label{app:cgamma}
  \hat C \hat\gamma_A &=\hat\gamma^T_A \hat C,\\
\label{app:cs}
  \hat C \hat S_{AB} &= -\hat S^T_{AB} \hat C.
\end{align} 

One can easily derive the matrix identity
\begin{equation}
\label{app:matid}
  \delta^\alpha_\gamma \delta^\beta_\delta = -\frac14 \hat C_{\gamma\delta}
  (\hat C^{-1})^{\alpha\beta} -\frac14 (\hat C \hat\gamma_A)_{\gamma\delta}
  (\hat\gamma^A\hat C^{-1})^{\alpha\beta}
  -\frac12 (\hat C \hat S_{AB})_{\gamma\delta}
  (\hat S^{AB} \hat C^{-1})^{\alpha\beta},
\end{equation}
which leads to various Fierz identities. Moreover, one
can simplify products involving two or more identical spinor factors. In
particular, one finds
\begin{align}
\label{app:id1}
  (\hth\hat\eta)(\hth\hat\varepsilon) &=
  -\frac14 (\hth\hth)(\hat\eta\hat\varepsilon)
  -\frac14(\hth\hat\gamma_A\hth)
  (\hat\eta\hat\gamma^A\hat\varepsilon),\\
\label{app:id2}
  (\hth\hat\eta)(\hth\hat\gamma_A\hth) &=
  - (\hth\hth)(\hat\eta\hat\gamma_A\hth).
\end{align}

\section{Calculation of the Killing Spinor} 
\label{app:calc}
In this appendix we shall calculate the matrix $\Lambda(x)$,
which relates Lorentz and $SO(4,1)$ spinors with each other. It will turn
out that the Lorentz spinor derived from the $SO(d+1,1)$ spinor
$\hth$ automatically is a Killing spinor.

According to equation \eqref{sup:thetasymm} $\hth$ is an
$SO(4,1)$ spinor. 
However, a spinor field $\psi(x)$ conventionally is a
Lorentz spinor, i.e.\ it transforms as a spinor under rotations of the local
Lorentz frame. Hence, we introduce a matrix $\Lambda(x)$ and demand
that the product $\hth\Lambda(x)\psi(x)$ be a scalar with
respect to symmetry transformations. 
The matrix $\Lambda(x)$
can be calculated using the knowledge of the transformation laws under the
$SO(4,1)$ symmetries. Thus,
\begin{align}
\notag
  \delta \left(\hth\Lambda(x)\psi(x)\right) &=
  - \delta\hth_\alpha \Lambda(x)^\alpha_{\;\beta} \psi(x)^\beta
  - \delta x^\mu \hth_\alpha \partial_\mu \left(
  \Lambda(x)^\alpha_{\;\beta} \psi(x)^\beta \right) \\
\label{app:ldef}
  &\equiv \hth_\alpha \Lambda(x)^\alpha_{\;\beta}
  \delta\psi(x)^\beta,
\end{align}
where $\delta\hth$ and $\delta x^\mu$ are given by equations
\eqref{sup:thetasymm} and \eqref{symm:xsymm}, respectively, and 
\[ \delta \psi = - \delta x^\mu D_\mu \psi + \frac12 D^\nu \delta
x^\mu \Sigma_{\mu\nu} \psi \]
As the parameters $a_i$, $b_i$, $\lambda$ and $\omega_{ij}$ in
$\delta x$ are independent, equation \eqref{app:ldef} yields the
following system of equations for $\Lambda$:
\begin{equation}
\label{app:lsys}
\begin{aligned}
  \hat S(P_i) \Lambda + \partial_i \Lambda &= 0, \\
  \hat S(D) \Lambda + x^\mu \partial_\mu \Lambda &=0, \\
  \hat S(K_i) \Lambda - 2 x^\mu \Lambda S_{i\mu} + \left( 2x_i
  x^\mu \partial_\mu -x^2 \partial_i \right) \Lambda &=0, \\
  \hat S(L_{ij}) \Lambda - \Lambda S_{ij} + \left(x_i\partial_j
  -x_j\partial_i \right) \Lambda &=0.
\end{aligned}
\end{equation}
The solution of equations \eqref{app:lsys} is not unique, but any
solution will suffice. A solution of equations \eqref{app:lsys} is  
\begin{align}
\label{app:lsol}
  \Lambda(x) &= \frac{\sqrt{x_0}}2 (1+\gamma_0) -
  \frac1{2\sqrt{x_0}} (1-\gamma_0)
  +\frac{x^i}{2\sqrt{x_0}} \gamma_i (1-\gamma_0) .\\
\intertext{It is also useful to know the inverse $\Lambda^{-1}$, which is
easily found to be} \label{app:linv}
  \Lambda^{-1}(x) &= \frac1{2\sqrt{x_0}} (1+\gamma_0)
  - \frac{\sqrt{x_0}}2 (1-\gamma_0)
  +\frac{x^i}{2\sqrt{x_0}} \gamma_i (1-\gamma_0).
\end{align}

Consider the spinor $(\hth\Lambda)_\alpha$, which by construction is a
Lorentz spinor. Since
$\hth\hat\eta=\hth\Lambda\Lambda^{-1}\hat\eta$, one finds
\begin{equation}
\label{app:cdef}
 (\Lambda^{-1}\hth)^\alpha = (C^{-1})^{\alpha\beta}
 (\hth \Lambda)_\beta,
\end{equation}
where $C^{-1}$ is the inverse of the charge conjugation matrix $C$ for Lorentz
spinors. Equation \eqref{app:cdef} yields $C = \Lambda^T \hat C \Lambda$,
which, together with equation \eqref{app:lsol}, leads to $C = -\hat C$.

Finally, one can check explicitly from the expression \eqref{app:lsol} that
\begin{equation}
\label{app:kill}
  D_\mu (\hth\Lambda)_\alpha = \frac12
  (\hth\Lambda\Gamma_\mu)_\alpha,
\end{equation}
which shows that $\theta(x) = \Lambda^{-1}(x)\hth$ is a Killing
spinor.
\end{appendix}


\begin{thebibliography}{99}
\bibitem{Maldacena} J.~Maldacena, Adv.~Theor.~Math.~Phys.~\textbf{2},
  231 (1998), \texttt{hep-th/9711200}  
\bibitem{Girardello} L.~Girardello, M.~Petrini, M.~Porrati and
  A.~Zaffaroni, JHEP \textbf{9812}, 022 (1998), \texttt{hep-th/9810126}
\bibitem{Gubser} S.~S.~Gubser, I.~R.~Klebanov and A.~M.~Polyakov,
  Phys.~Lett.~\textbf{B428}, 105 (1998), \texttt{hep-th/9802109}
\bibitem{Witten} E.~Witten, Adv.~Theor.~Math.~Phys.~\textbf{2}, 253
  (1998), \texttt{hep-th/9802150}
\bibitem{Ferrara3} S.~Ferrara and A.~Zaffaroni, \emph{Bulk Gauge
  Fields in AdS Supergravity and Supersingletons}, \texttt{hep-th/9807090}
\bibitem{Mueck1} W.~M\"uck and K.~S.~Viswanathan, Phys.~Rev.~D
  \textbf{58}, 041901 (1998), \texttt{hep-th/9804035}
\bibitem{Freedman} D.~Z.~Freedman, S.~D.~Mathur, A.~Matusis and
  L.~Rastelli, \emph{Correlation functions in the CFT(d)/AdS(d+1)
  correspondence}, \texttt{hep-th/9804058}
\bibitem{Henningson} M.~Henningson and K.~Sfetsos,
  Phys.~Lett.~\textbf{B431}, 63 (1998), \texttt{hep-th9803251} 
\bibitem{Mueck2} W.~M\"uck and K.~S.~Viswanathan, Phys.~Rev.~D
  \textbf{58}, 106006 (1998), \texttt{hep-th/9805145}
\bibitem{Minces} P.~Minces and V.~O.~Rivelles, \emph{Chern-Simons
  theories in the AdS/CFT correspondence}, \texttt{hep-th/9902123}
\bibitem{Liu} H.~Liu and A.~A.~Tseytlin, Nucl.~Phys.~B \textbf{533},
  88 (1998), \texttt{hep-th/9804083}
\bibitem{Arutyunov} G.~E.~Arutyunov and S.~A.~Frolov, \emph{On the
  origin of  supergravity boundary terms in the AdS/CFT
  correspondence}, \texttt{hep-th/9806216}
\bibitem{Mueck3} W.~M\"uck and K.~S.~Viswanathan, \emph{The graviton
  in the AdS-CFT correspondence: Solution via the Dirichlet boundary
  value problem}, \texttt{hep-th/9810151}
\bibitem{Corley} S.~Corley, Phys.~Rev.~D \textbf{59}, 086003 (1998), 
  \texttt{hep-th/9808184}
\bibitem{Volovich} A.~Volovich, JHEP \textbf{9809}, 022 (1998),
  \texttt{hep-th/9809009} 
\bibitem{Koshelev} A.~S.~Koshelev and O.~A.~Rytchkov, \emph{Note on
  the massive Rarita-Schwinger field in the AdS/CFT correspondence},
  \texttt{hep-th/9812238}
\bibitem{Ferrara1} S.~Ferrara and A.~Zaffaroni,
  Phys.~Lett.~\textbf{B431}, 49 (1998), \texttt{hep-th/9803060}
\bibitem{Ferrara2} S.~Ferrara, A.~Kehagias, H.~Partouche and
  A.~Zaffaroni,  Phys.~Lett.~\textbf{B431}, 57 (1998),
  \texttt{hep-th/9804006}
\bibitem{Banados} M.~Banados, K.~Bautier, O.~Coussaert, M.~Henneaux
  and M.~Ortiz, Phys.~Rev.~D \textbf{58}, 085020 (1998)
\bibitem{Nishimura} M.~Nishimura and Y.~Tanii, \emph{Supersymmetry in
  the AdS/CFT Correspondence}, \texttt{hep-th/9810148}
\bibitem{Gunaydin} M.~G\"unaydin, D.~Minic and M.~Zagermann,
  \emph{Novel supermultiplets of $SU(2,2|4)$ and the AdS$_5$/CFT$_4$
  duality}, \texttt{hep-th/9810226}
\bibitem{Ito} K.~Ito \emph{Extended Superconformal Algebras on
  AdS$_3$}, \texttt{hep-th/9811002}
\bibitem{Anselmi} D.~Anselmi and A.~Kehagias, \emph{Subleading
  Corrections and Central Charges in the AdS/CFT Correspondence},
  \texttt{9812092} 
\bibitem{Figueroa} J.~M.~Figueroa-O'Farrill, \emph{On the
  supersymmetries of anti de Sitter vacua}, \texttt{hep-th/9902066}
\bibitem{Yamaguchi} S.~Yamaguchi, Y.~Ishimoto and K.~Sugiyama,
  \emph{AdS$_3$/CFT$_2$ Correspondence and Space-Time $\mathcal{N}=3$ 
  Superconformal Algebra}, \texttt{hep-th/9902079} 
\bibitem{Breitenlohner} P.~Breitenlohner and D.~Z.~Freedman,
  Ann.~Phys.~\textbf{144}, 249 (1982)
\bibitem{Burges} C.~J.~C.~Burges, D.~Z.~Freedman, S.~Davies and
  G.~W.~Gibbons, Ann.~Phys.~\textbf{167}, 285 (1986)
\bibitem{Weinberg} S.~Weinberg, \emph{Gravitation and
  Cosmology: Principles and Applications of the General Theory of
  Relativity}, J.~Wiley \& Sons (1972)
\bibitem{vanHolten} J.~W.~van~Holten and A.~Van~Proeyen, J.~Phys.~A
  \textbf{15}, 3763 (1982)
\bibitem{Cornwell3} J.~F.~Cornwell, \emph{Group Theory in Physics},
  vol.~3, Academic Press (1989)
\bibitem{diFrancesco} P.~DiFrancesco, P.~Mathieu and D.~S\'en\'echal,
  \emph{Conformal Field Theory}, Springer Verlag (1996)
\bibitem{Zumino} B. Zumino, Nucl.~Phys.~B \textbf{127}, 189 (1977)
\bibitem{Keck} B.~W.~Keck, J.~Phys.~A \textbf{8}, 1819 (1975)
\bibitem{Ivanov} E.~A.~Ivanov and A.~S.~Sorin, J.~Phys.~A
  \textbf{13}, 1159 (1980)
\bibitem{Henneaux} M.~Henneaux, \emph{Boundary terms in the AdS/CFT
  correspondence for spinor fields}, \texttt{hep-th/9902137}
\bibitem{Pilch} K.~Pilch, P.~van~Nieuwenhuizen and M.~F.~Sohnius,
  Commun.~Math.~Phys.~\textbf{98}, 105 (1985)
\end{thebibliography}
\end{document}